\documentstyle[epsfig]{aipproc}

\catcode`@=11
\def\citer{\@ifnextchar[{\@tempswatrue\@citexr}{\@tempswafalse\@citexr[]}}

\def\@citexr[#1]#2{\if@filesw\immediate\write\@auxout{\string\citation{#2}}\fi
  \def\@citea{}\@cite{\@for\@citeb:=#2\do
    {\@citea\def\@citea{--\penalty\@m}\@ifundefined
       {b@\@citeb}{{\bf ?}\@warning
       {Citation `\@citeb' on page \thepage \space undefined}}%
\hbox{\csname b@\@citeb\endcsname}}}{#1}}
\catcode`@=12

\newcommand{\beq}{\begin{eqnarray}}
\newcommand{\eeq}{\end{eqnarray}}

\newcommand{\s}{\\ \vspace*{-3.5mm} }

\begin{document}
\hspace{0.9cm}\rightline{
        \begin{minipage}{4cm}
        PM/01--04\\
        hep-ph/0101263\hfill \\
        \end{minipage}        
}

\title{Production of MSSM Higgs Bosons at Future $\gamma\gamma$ Colliders}
\author{M.M.~M\"uhlleitner}
\address{Laboratoire de Physique Math\'ematique et Th\'eorique, 
UMR5825--CNRS, Universit\'e de Montpellier II, F--34095 Montpellier Cedex 5,
France}
\maketitle
\vspace{-0.55cm}

\begin{abstract}
Future $\gamma\gamma$ colliders allow the production of the heavy neutral MSSM Higgs bosons $H$ and $A$ as single resonances. The prospects of finding these particles in the $b\bar{b}$ and the neutralino-pair final states have been analysed. The $H,A$ bosons can be discovered for medium values of $\tan\beta$ with masses up to 70--80\% of the initial $e^\pm e^-$ c.m.\ energy. This production mode thus covers parts of the supersymmetric parameter space that are not accessible at other colliders.
%
\end{abstract}
\vspace{-0.5cm}

{\bf 1.} The discovery of Higgs bosons \cite{higgs} is the {\it experimentum crucis} of the electroweak sector of the Standard Model and its supersymmetric extensions. The minimal supersymmetric model (MSSM) includes five physical Higgs states \cite{habergun}, two neutral CP-even ($h,H$), a CP-odd ($A$) and two charged ($H^\pm$) scalars. The light scalar Higgs boson $h$ with a theoretical upper mass bound of about $130$~GeV \cite{mssmrad} can be found at existing and future $p\bar{p}/pp$ and $e^+e^-$ colliders. Apart from a small region in the supersymmetric parameter space, the heavy states $H,A$, however, might escape discovery at the LHC for moderate values of $\tan\beta$ and masses beyond $\sim 200$~GeV \cite{lhc}. At 500 GeV $e^+ e^-$ linear colliders, where heavy MSSM Higgs bosons are produced in the associated channel \cite{3b}, $e^+ e^- \to HA$, they can be discovered only for masses below $\sim 250$~GeV. 

Photons produced by Compton scattering off the incoming $e^\pm e^-$ beams can reach c.m.~energies of 70--80\% of the initial collider energies and high degrees of polarization \cite{plc}. By resonant production of the neutral Higgs bosons \cite{borden}, $\gamma\gamma \to h,H,A$,
future $\gamma\gamma$ colliders extend the Higgs discovery to higher mass regions. With integrated annual luminosities of 300~fb$^{-1}$ \cite{9a}, sufficiently high signal rates can be achieved so that photon colliders provide a powerful instrument for the search of Higgs bosons in regions of the parameter space not accessible elsewhere \cite{muehldiss}. \s

\begin{figure}[hbtp]
\vspace{-0.5cm}
\begin{center}
\epsfig{figure=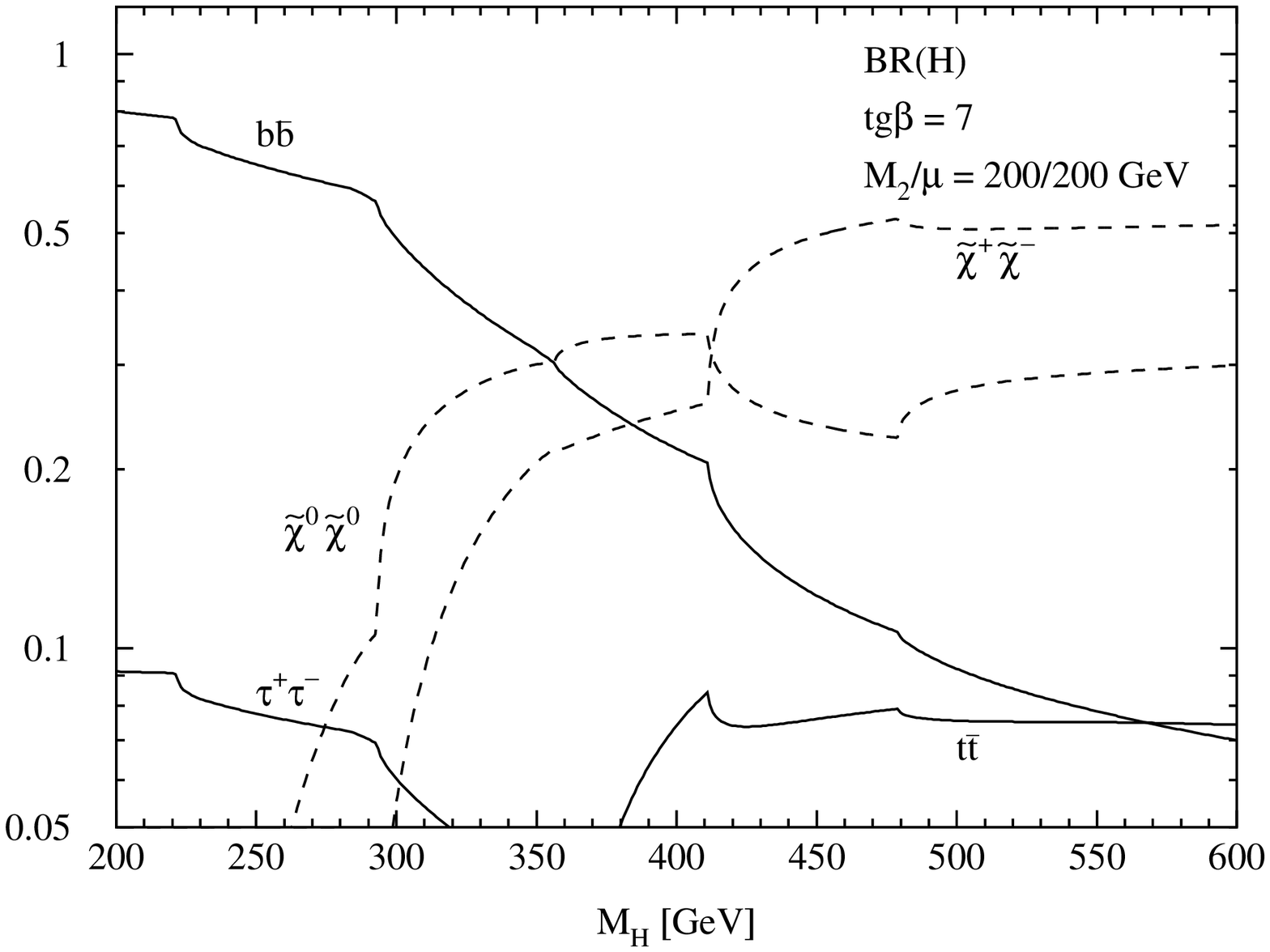,width=7cm}
\hspace{0mm}
\epsfig{figure=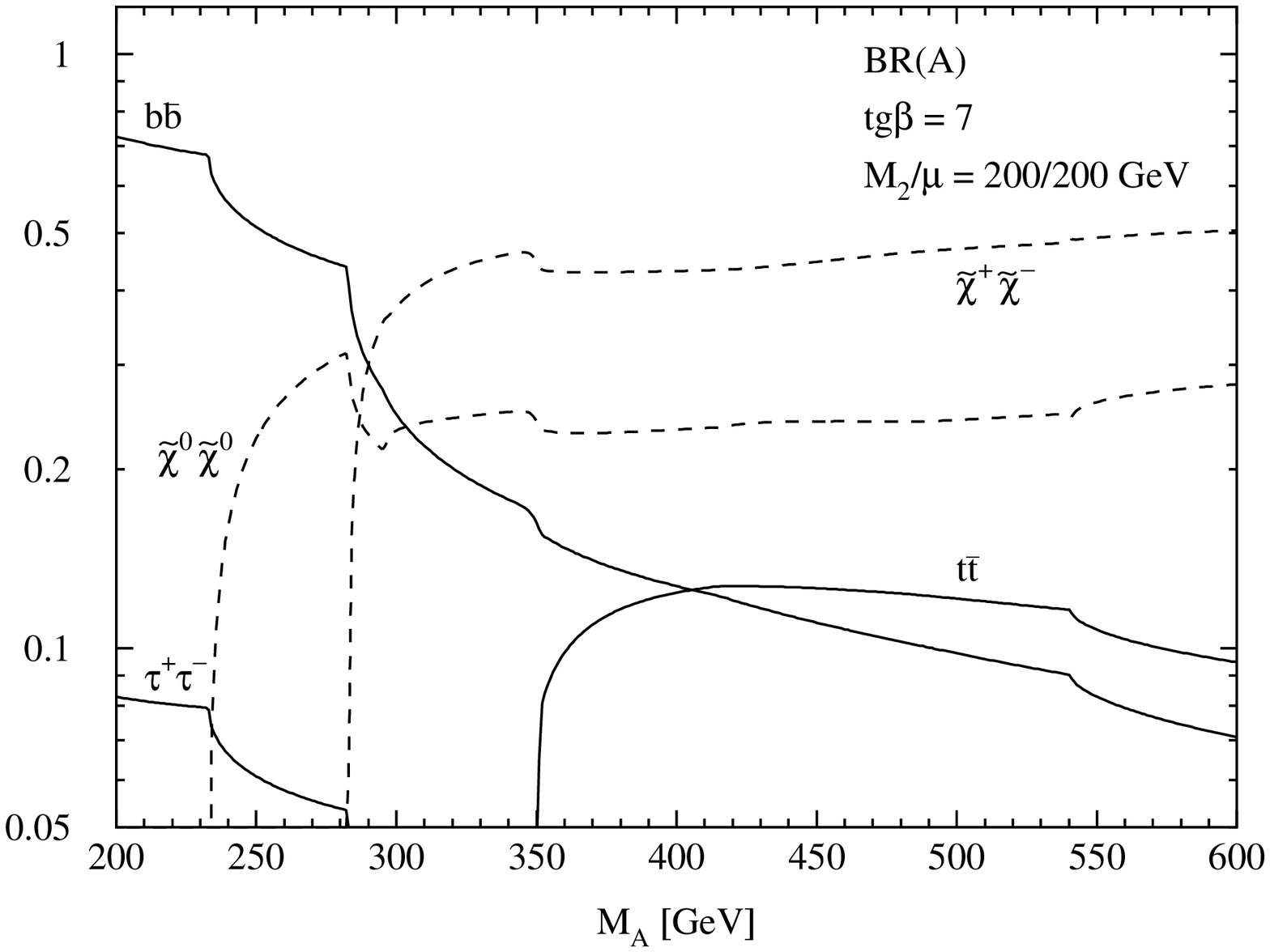,width=7cm}
\end{center}
\caption[]{\label{fig:br} Branching ratios of the heavy Higgs bosons
$H,A$ as a function of the corresponding Higgs mass. $\tilde{\chi}\tilde{\chi}$ represents the sum of all supersymmetric particles except the LSP pair. The gaugino and neutralino masses are $m_{\tilde{\chi}_{1,2}^\pm}=141,270$~GeV, $m_{\tilde{\chi}_{1,2,3,4}^0}=85,148,208,271$~GeV.}
\end{figure}

{\bf 2.} We have analysed the $H$ and $A$ production in their most promising decay channels, $H,A\to b\bar{b}$ and $\tilde{\chi}_0 \tilde{\chi}_0$ \cite{muehldiss}. Fig.~\ref{fig:br} shows the branching ratios of $H,A$ \citer{41,43} for $\tan\beta=7$ and $M_{H,A}>200$~GeV, a parameter region not covered by LHC discovery modes. The higgsino and gaugino MSSM parameters have been set $\mu=M_2=200$~GeV with a universal gaugino mass at the GUT scale. Squarks and sleptons are assumed to be so heavy that they do not affect the results significantly.
\vspace{-0.2cm}

Due to the enhancement of the MSSM Higgs boson couplings to $b\bar{b}$ at large $\tan\beta$, the $b\bar{b}$ decay is sizeable and dominates for moderate masses. Beyond the corresponding thresholds, the decays into neutralino and chargino pairs become the most important channels. The branching ratios into $\tau^+\tau^-$ and into $t\bar{t}$ pairs are of ${\cal O}(10\%)$ or less. \s

{\bf 3.} $b\bar{b}$ {\bf Channel.} Fig.~2a shows the result for $\gamma\gamma \to b{\bar b}$ and polarized $e^\pm e^-$ and photon beams. 
%
%
Assuming that evidence for a Higgs boson has been found in a preliminary
rough scan of the $\gamma\gamma$ energy, we choose the maximum of the
$\gamma\gamma$ luminosity spectrum to lie at the mass $M_A$. 
This maximum for equal photon helicities is at 70--80\% of the 
$e^\pm e^-$ c.m. energy \cite{plc,kuehn}.
 A cut $\Delta=\pm 3$~GeV \cite{schreiber} of the photon spectrum around $\sqrt{s_{\gamma\gamma}}=M_A$ accounts for the limited energy resolution in the final state. The NLO QCD corrections to the signal \cite{42,hggqcd}, background \cite{bkgqcd} and interference term \cite{muehldiss} are included. In order to increase the significance of the signal, the final states have been restricted to the two-jet configurations [for details, see Ref.~\cite{muehldiss}]. Higher order corrections that become important in the two-jet final states have been taken into account by (non-)Sudakov form factors \cite{resum}. A cut in the production angle $\theta$ of the $b$-quarks leads to a further enhancement of the significance of the signal.

As can be inferred from Fig.~2a, heavy Higgs bosons can be discovered up to about 70--80\% of $\sqrt{s}_{ee}$. Thus, at a 500~GeV collider, scalars with masses up to $\sim 400$~GeV can be found, while above a c.m.\ energy of 800 GeV the mass reach can be extended to about 600~GeV before the signal becomes too small for detection.

Due to the finite experimental energy resolution both resonances $H$ and $A$ are included in the signal peak [$M_H - M_A\sim 1$~GeV]. Fig.~2b shows that at least in parts of the supersymmetric parameter space the two resonances can be separated by performing a fine scan of the resonance region with $\Delta = \pm 2$~GeV. 
\begin{figure}[ht]
\begin{center}
\epsfig{figure=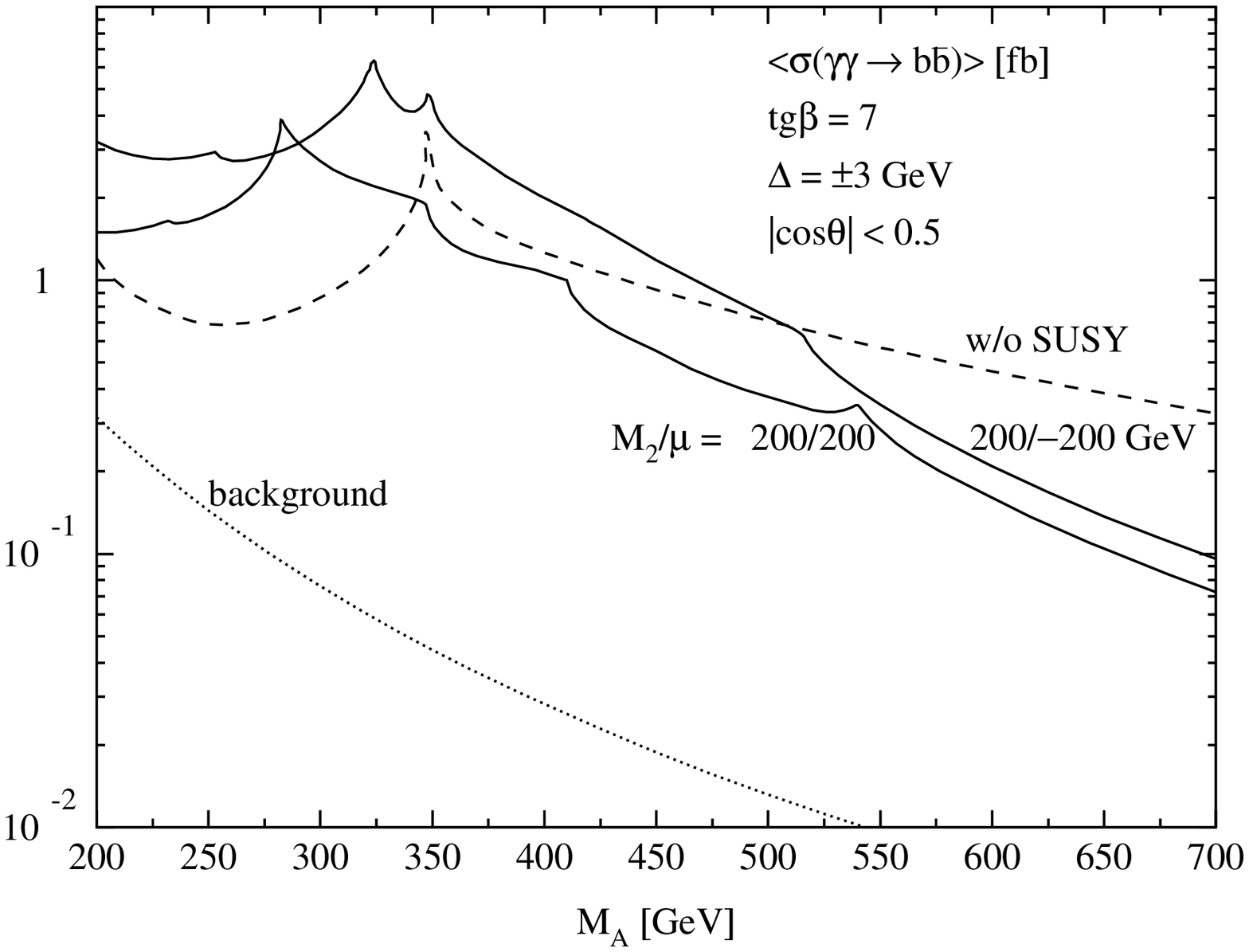,width=6cm}
\hspace{0mm}
\epsfig{figure=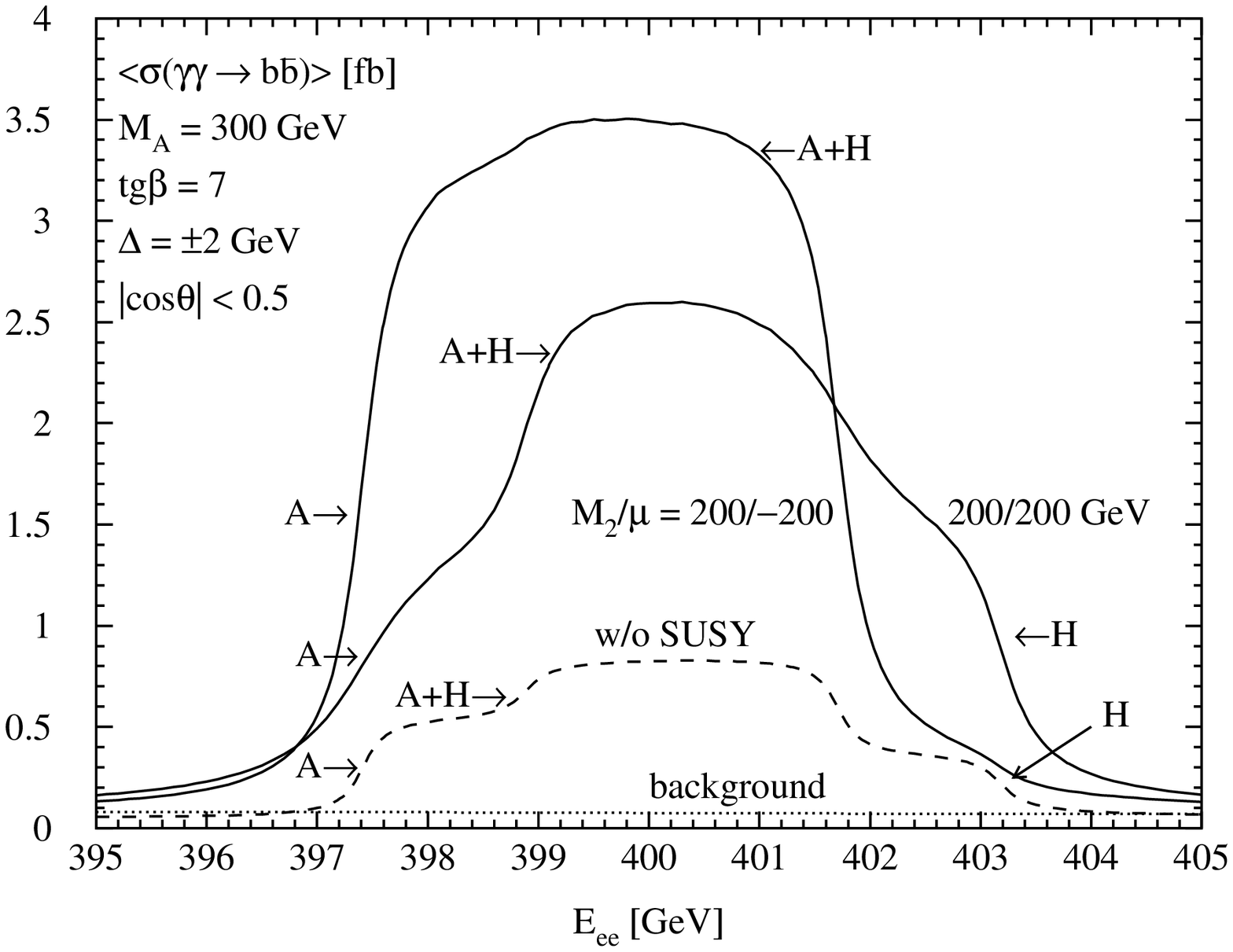,width=6cm}
\end{center}
\caption[]{(a) Cross sections for the resonant $H,A$ production in $\gamma\gamma$ fusion as a function of $M_A$ with final decays into $b\bar{b}$, and the corresponding background cross section. The SUSY parameters are chosen $\tan\beta=7$, $M_2=\pm\mu=200$~GeV. For comparison the signal cross section in the case of vanishing SUSY-particle contributions is also shown. (b) Threshold scans for $H,A$ production as a function of the $e^\pm e^-$ c.m.~energy for $b\bar{b}$ final states.}
\label{fig:bb}
\end{figure}
\vspace*{-0.6cm}

\begin{figure}[hbtp]
\begin{center}
\epsfig{figure=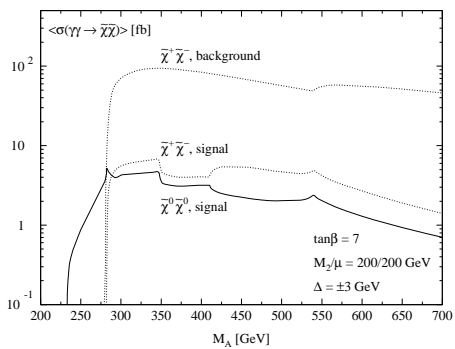,width=6cm}
\end{center}
\caption[]{Same as Fig.~2a, but for chargino and neutralino final states.}
\label{fig:susy}
\end{figure}

{\bf 4.} $\tilde{\chi}_0\tilde{\chi}_0$ {\bf Channels.} The significant branching ratios into SUSY particles, cf.~Figs.~\ref{fig:br}, lead to quite sizeable signal cross sections for chargino and neutralino pair final states, as can be inferred from Fig.~\ref{fig:susy}. Because of the integer chargino charge the background cross section, however, is an order of magnitude larger. At leading order neutralino pair production is not possible in $\gamma\gamma$ collisions so that this channel provides an additional mode for the $H,A$ discovery \cite{muehldiss,belan}. Below the chargino threshold this is evident. As soon as the chargino channel opens, the different topologies of neutralino and chargino cascade decays have to be exploited in order to separate the neutralino signal from the background charginos.
\vspace{-0.05cm}

{\bf 5.} {\bf In summary.} The neutral heavy Higgs bosons of the MSSM, $H,A$, can be discovered for medium values of $\tan\beta$ at future $\gamma\gamma$ colliders with masses up to 400~GeV at a c.m.~energy of 500~GeV of the initial $e^\pm e^-$ beams. At a TeV collider the mass reach can be extended beyond 600~GeV. The Higgs discovery via $\gamma\gamma$ fusion thus enters parameter regions neither accessible in the respective $e^\pm e^-$ mode of linear colliders nor at the LHC.\s

{\bf Acknowledgements.} I am grateful to my collaborators M.~Kr\"amer, M.~Spira and P.M.~Zerwas for fruitful discussions. I thank the organizers of the Linear Collider Workshop at Fermilab Oct 2000 for the stimulating atmosphere. This work is supported by the European Union under contract HPRN-CT-2000-00149.

\end{document}